\documentclass[a4paper,10pt]{article}
\usepackage[utf8]{inputenc}
\usepackage{authblk}

\usepackage{graphicx}
\usepackage{epstopdf}
\usepackage{todonotes}
\usepackage{lineno,hyperref}
\usepackage{upgreek}
\usepackage{units}
\usepackage{multirow}
\usepackage{amssymb}	
\usepackage{amsmath}	
\usepackage{textcomp}
\usepackage{pdfpages}

\usepackage{caption}
\usepackage{subcaption}

\title{Investigations into the impact of locally modified sensor architectures on the detection efficiency of silicon micro-strip sensors}

\author[1]{Luise Poley}
\author[1]{Kristin Lohwasser}
\author[2]{Andrew Blue}
\author[3]{Mathieu Benoit}
\author[1]{Ingo Bloch}
\author[1]{Sergio D\'{i}ez}
\author[4]{Vitaliy Fadeyev}
\author[5]{Bruce Gallop}
\author[6]{Ashley Greenall}
\author[1]{Ingrid-Maria Gregor}
\author[1]{John Keller}
\author[7]{Carlos Lacasta}
\author[2]{Dzmitry Maneuski}
\author[3,6]{Lingxin Meng}
\author[1]{Marko Milovanovic}
\author[8]{Ian Pape}
\author[5]{Peter W. Phillips}
\author[9]{Laura Rehnisch}
\author[8]{Kawal Sawhney}
\author[5]{Craig Sawyer}
\author[9]{Dennis Sperlich}
\author[1]{Martin Stegler}
\author[10]{Yoshinobu Unno}
\author[11]{Matt Warren}
\author[1]{Eda Yildirim}

\affil[1]{Deutsches Elektronen-Synchrotron, Notkestra{\ss}e, Hamburg, Germany}
\affil[2]{SUPA School of Physics and Astronomy, University of Glasgow, University Avenue, Glasgow, United Kingdom}
\affil[3]{D\'{e}partement de Physique Nucl\'{e}aire et Corpusculaire, University of Geneva, quai Ernest Ansermet, Gen\`{e}ve, Suisse}
\affil[4]{Santa Cruz Institute of Particle Physics, University of California, High Street, Santa Cruz, United States of America}
\affil[5]{Particle Physics Department, STFC Rutherford Appleton Laboratory, Harwell Science and Innovation Campus, Didcot, United Kingdom}
\affil[6]{Department of Physics, University of Liverpool, Cambridge Street, Liverpool, United Kingdom}
\affil[7]{Instituto de F\'{\i}sica Corpuscular, CSIC-U. Valencia, c/ Catedr\'{a}tico Jos\'{e} Beltr\'{a}n, Paterna, Spain}
\affil[8]{Diamond Light Source Ltd, Diamond House, Harwell Science and Innovation Campus, Didcot, United Kingdom}
\affil[9]{Institut für Physik, Humboldt-Universit\"{a}t zu Berlin, Newtonstra{\ss}e, Berlin, Germany}
\affil[10]{Institute of Particle and Nuclear Study, KEK, Oho, Tsukuba, Japan}
\affil[11]{Department of Physics and Astronomy, University College London, Gower Street, London, United Kingdom}

\begin{document}

\maketitle

\begin{abstract}
The High Luminosity Upgrade of the LHC will require the replacement of the Inner Detector of ATLAS with the Inner Tracker (ITk) in order to cope with higher radiation levels and higher track densities. Prototype silicon strip detector modules are currently developed and their performance is studied in both particle test beams and X-ray beams. In previous test beam studies of prototype modules, the response of silicon sensors has been studied in detailed scans across individual sensor strips. These studies found instances of sensor strips collecting charge across areas on the sensor deviating from the geometrical width of a sensor strip. The variations have been linked to local features of the sensor architecture.

This paper presents results of detailed sensor measurements in both \mbox{X-ray} and particle beams investigating the impact of sensor features (metal pads and \mbox{p-stops}) on the sensor strip response.
\end{abstract}

\section{Introduction} 

In the current layout for silicon strip sensor modules for the future ATLAS Inner Tracker, modules consist of silicon strip sensors, printed circuit boards (hybrids)~\cite{31_hybsandmods} and binary readout chips (ABC130 ASICs~\cite{ABC130}). Readout chips are glued on to hybrids, which are then glued on to sensors. Electrical connections between readout chips and hybrids are made by aluminium wire bonds (diameter \unit[25]{$\upmu$m}), providing both power for the chips and data readout connections. 

\begin{figure}
\centering
\includegraphics[width=0.5\textwidth]{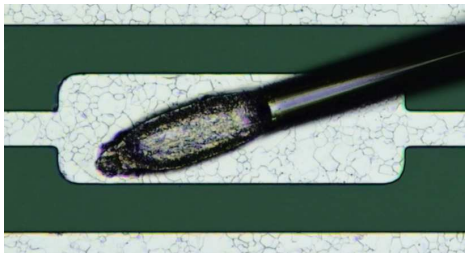}
\caption{Laser microscope image of a wire bond foot on an aluminium bond pad on a silicon sensor.}
\label{fig:8_bondfuss}
\end{figure}

Wire bonds also connect each sensor strip to an ASIC readout channel: the energy deposited in the bulk by a traversing charged particle or absorbed photon is detected in 1-2 silicon strips, providing spatial information corresponding to the pitch of a sensor strip (\unit[74.5]{$\upmu$m}). Each sensor strip is read out individually by one ASIC channel. The connection of ASICs and sensor strips by wire bonds requires the addition of electrically conductive bond pads to the aluminium layer on top of each strip implant. The dimensions of these bond pads are defined by necessities for safe wire bonding:
\begin{itemize}
 \item a single bond foot (the area over which a wire bond is connected to a bond pad) has a width and length of up to \unit[$35\times120$]{$\upmu\text{m}^2$} (see figure~\ref{fig:8_bondfuss})
 \item a typical wire bonding wedge used for this application has a width of about \unit[80]{$\upmu$m}
 \item in case of wire bonding failures, further wire bonding attempts can be necessary, requiring a bond pad size sufficient to place two wire bond feet side by side
\end{itemize}
Consequently, bond pads were chosen to have an approximately rectangular shape of \unit[$56\times200$]{$\upmu\text{m}^2$} (see figure~\ref{fig:8_bondfuss}), i.e. close to the strip pitch (\unit[74.5]{$\upmu$m}).

\section{Sensors under investigation}

\begin{figure}
\centering
\includegraphics[width=0.5\textwidth]{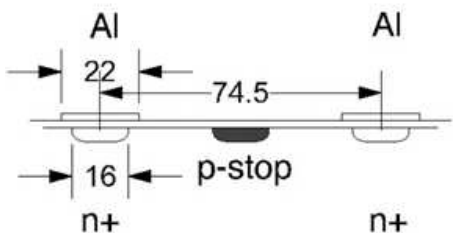}
\caption{Layout of the silicon strip sensors studied in test beam measurements: adjacent n$^+$-doped strip implants (strip pitch {\unit[74.5]{$\upmu$m}}) in a p-doped bulk are electrically separated by p-doped implants (p-stops)~\cite{8_doping}.}
\label{fig:8_slayout}
\end{figure}

Silicon strip sensors for the ATLAS Inner Tracker consist of a p-doped bulk with n-doped strip implants~\cite{8_doping}. In order to isolate individual strip implants, p-doped implants (p-stops)~\cite{ref_pstops} are positioned between strip implants. Figure~\ref{fig:8_slayout} shows the layout of the sensors used for the measurements presented here.

\begin{figure}
\centering
\includegraphics[width=\textwidth]{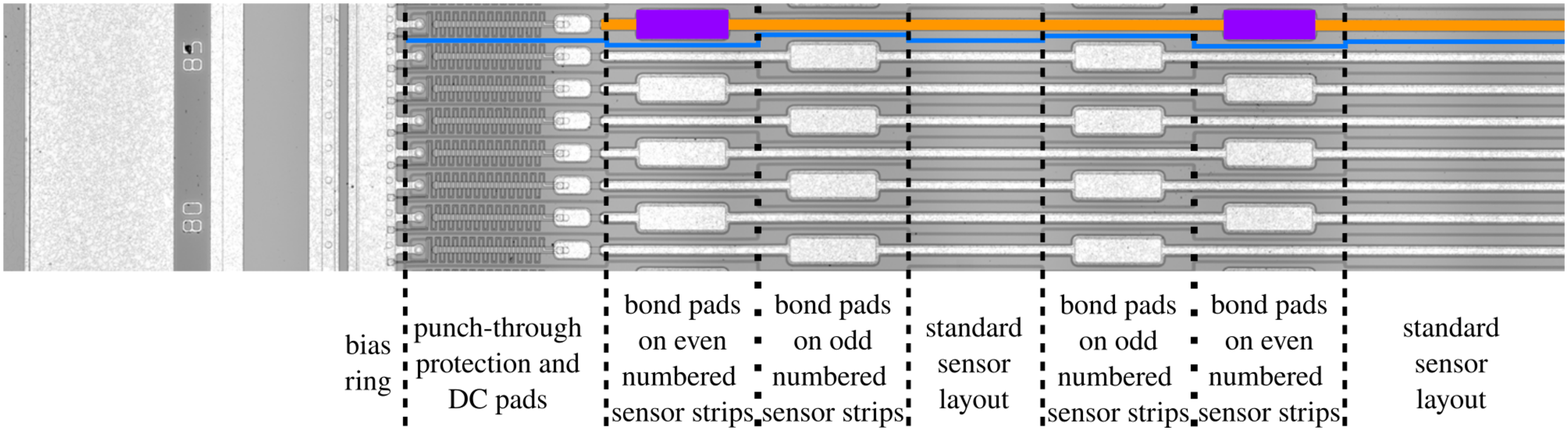}
\caption{Laser microscope image of a miniature sensor showing bias ring, punch-through protection~\cite{PTP} and sensor strips (orange) with bond pads (violet), electrically separated by p-stop implants (blue). One complete row of bond pads, comprised of one row of bond pads on even numbered sensor strips and one row of bond pads on odd numbered strips, stretches over a length of {\unit[700]{$\upmu$m}.}}
\label{fig:8_1}
\end{figure}

AC-coupled bond pads are added to the metal layer on top of strip implants (see figure~\ref{fig:8_1}) to allow wire bonding. With a strip pitch of \unit[74.5]{$\upmu$m} and a bond pad width of \unit[56]{$\upmu$m} for safe wire bonding, bond pads need to be positioned in a staggered design of two rows, alternating on odd and even numbered sensor strips. Below bond pads, the width of the strip implant is increased to cover the full bond pad area. P-stop implants, which for most of the length of strips are straight and at the centre between two neighbour sensor strip implants, are arranged around these bond pads, leading to uneven distances between p-stops and strip implants (see figure~\ref{fig:8_1}). 

ATLAS07~\cite{8_doping} and ATLAS12~\cite{8_ATLAS12} sensors were produced as realistic prototypes for sensor tests, with a sensor architecture similar to the sensors to be used in the future ATLAS strip tracker.

First indications that sensor strip responses might be different in sensor regions with and without bond pads were found in measurements aiming to investigate charge sharing between adjacent strips~\cite{8-Diamond1}. These measurements were conducted at beam line B16~\cite{B16} of the Diamond Light Source using a micro-focused X-ray beam with a photon energy of \unit[15]{keV} (see section~\ref{sec:Dia16}).
Scanning the bond pad area of adjacent silicon sensor strips in an X-ray beam, the responding width of individual sensor strips had been found to match the uneven p-stops positions in that area rather than the strip pitch.

\section{Particle test beam measurements}
\label{sec:DESYbeam}

Further measurements were performed using an ATLAS07 miniature sensor prototype ~\cite{8_doping} with an active area of $\unit[\sim7\times7]{\text{mm}^2}$. Test beam measurements were performed with the sensor placed inside a beam telescope (described in~\cite{7_EUDET}). The beam telescope consists of six planes of MIMOSA26 pixel sensors, arranged in groups of three in front of and three behind the device under test. A particle passing through the sensor under investigation is also registered in each telescope plane. MIMOSA26 sensor pixels have a pitch of \unit[18.4]{$\upmu$m}, allowing the reconstruction of each particle track with high spatial resolution ($\mathcal{O}(\upmu$m))~\cite{45_tel}. 

The tracks of the beam particles traversing the telescope are reconstructed from signal clusters reconstructed in the telescope planes using the General Broken Lines (GBL) Algorithm \cite{GBL}. The alignment parameters are calculated using the Millepede-II Algorithm \cite{60_mille}.

The charge deposited in the sensor was read out via wire bonds connecting the sensor to the analogue readout system ALiBaVa \cite{7-Alibava1}. The charge collected in each ALiBaVa readout channel is compared to the average expected noise in each channel. The collected charge in the channel with the largest signal-to-noise ratio (SNR) is used to form a cluster. Tracks where $\text{SNR} < 5$ are rejected. The charge collected in adjacent channels is added to the cluster if the SNR in the respective channel exceeds three. The reconstructed clusters are then taken to be hits caused by the traversing particles from the beam. Each hit found by the readout system can be related to a particle track reconstructed in the beam telescope from a given beam spill. By investigating which sensor strip responded with a signal and relating this to the expected hit position on the sensor given the parameters of the particle track reconstructed in the beam telescope, charge collection from individual strips can be mapped in the x-y plane.

Both the ALiBaVa daughterboard (used for signal readout) and the sensor board (holding the miniature sensors) were mounted on a copper plate cooled down to \unit[10]{\textcelsius}. The cooling plate was mounted inside a plastic housing to minimise light exposure. The sensor was operated fully depleted at a bias voltage of \unit[-250]{V}. 

\subsection{Results}

\begin{figure}
\centering
\includegraphics[width=\textwidth]{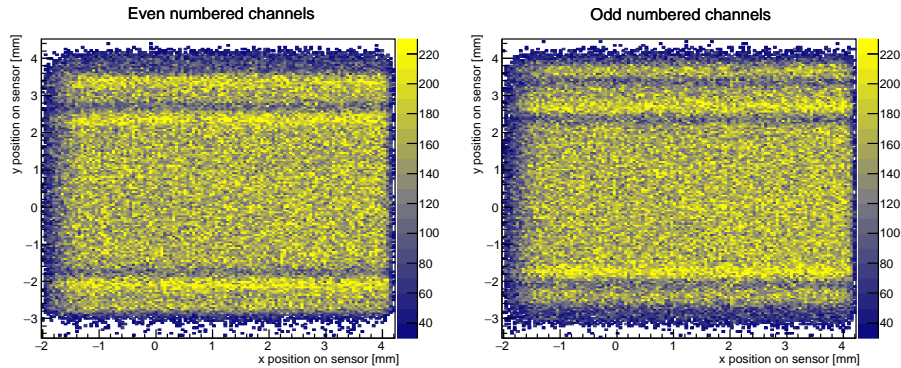}
\caption{Positions of particle hits on an ATLAS07 miniature sensor for \unit[6]{million} events. The sensor strip implants are oriented parallel to the y-axis. The left plot shows the hit map of only even numbered sensor strips, the right plot shows the hits collected only by odd numbered strips. Areas with fewer hits are paired with areas with more hits, matching the positions of bond pad rows consisting of one row of bond pads for even numbered channels and one row of bond pads for odd numbered channels each.}
\label{fig:8_DESYTB1}
\end{figure}

\begin{figure}
\centering
\includegraphics[width=0.99\textwidth]{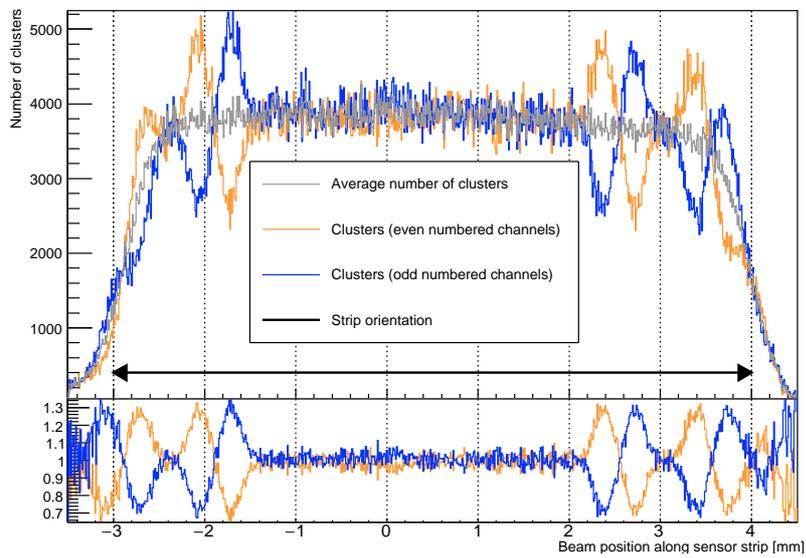}
\caption{{\unit[6]{million}} hits collected on an ATLAS07 miniature sensor, projected on on the sensor axis parallel to the strip implants. Separating the clusters according to the highest responding channel (odd/even) shows regions where the number of clusters in odd/even channels deviates from the average by up to {\unit[30]{\%}}.}
\label{fig:Vitaliy}
\end{figure}

Figure~\ref{fig:8_DESYTB1} shows the resulting map of clusters obtained for one ATLAS07 miniature sensor.
The number of recorded particle hits per sensor position showed that the presence of bond pads leads to a statistical effect on the number of recorded hits: sensor strips with bond pads show an increased number of hits in the bond pad area, while sensor strips without bond pads show fewer hits in the same area. Since the overall number of hits (i.e. the combined hits from odd and even numbered channels) is approximately constant over the whole sensor area, an increased/decreased number of hits indicates hits were collected over a larger/smaller sensor area. 
Figure~\ref{fig:Vitaliy} shows a projection of the number of collected hits in order to attempt a quantification of the effect.
After the previous test beam results had indicated that bond pads might lead to different widths over which a sensor strip responds, the findings from this test beam showed a similar effect: the presence of a bond pad on a sensor strip results in this strip collecting hits over a larger area than intended. This effect leads to an average difference of up to \unit[30]{\%} in number of collected clusters in bond pad regions compared to sensor regions without bond pads.

\begin{figure}
\centering
\begin{subfigure}{.49\textwidth}
  \centering
  \includegraphics[width=\linewidth]{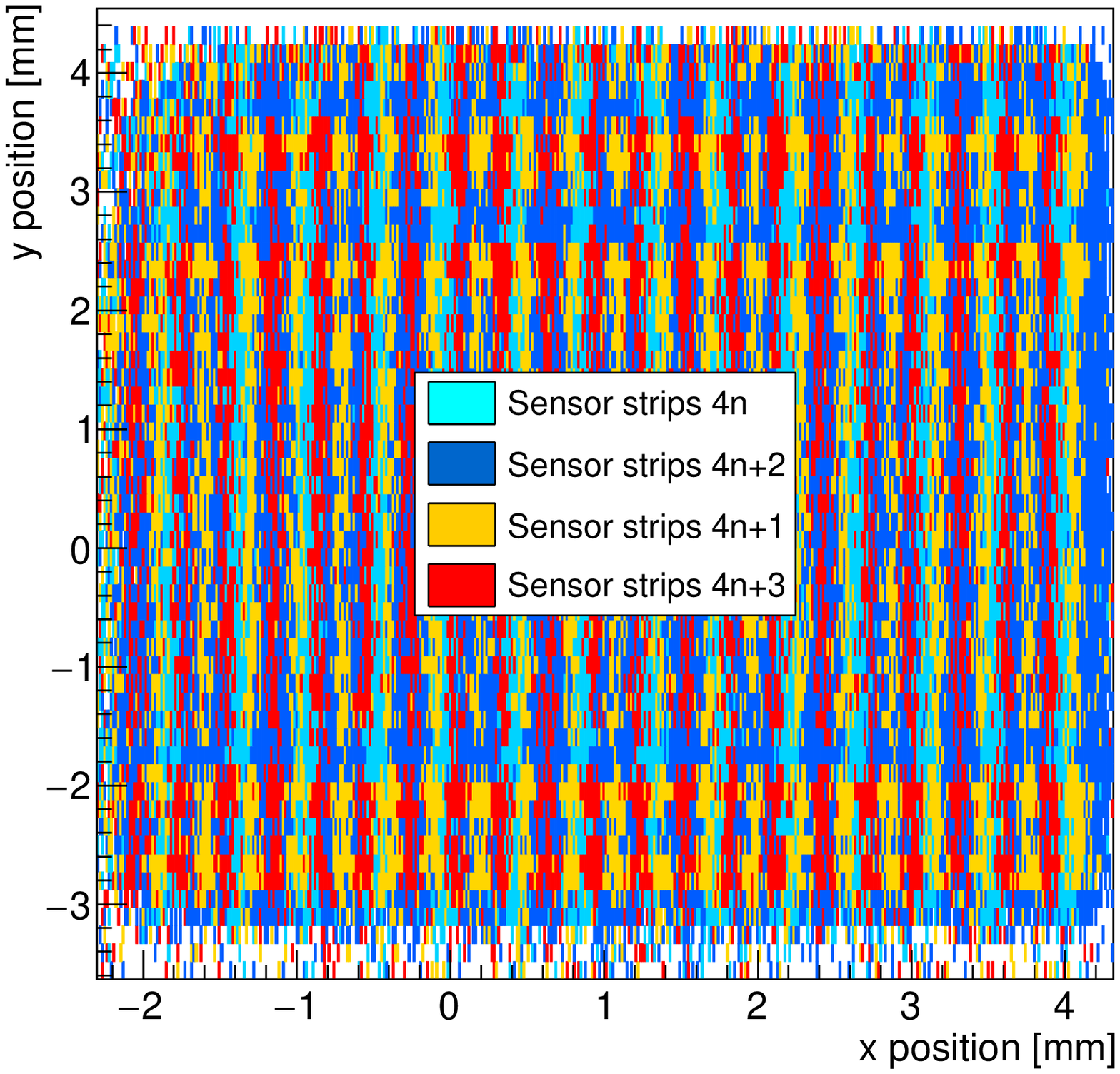}
  \caption{ATLAS07 miniature sensor map showing the mainly responding sensor strip for a given position on the sensor: orange and red bins represent odd numbered channels, cyan and blue bins represent even numbered channels.}
  \label{fig:8_maps_a}
\end{subfigure}
\begin{subfigure}{.46\textwidth}
  \centering
  \includegraphics[width=\linewidth]{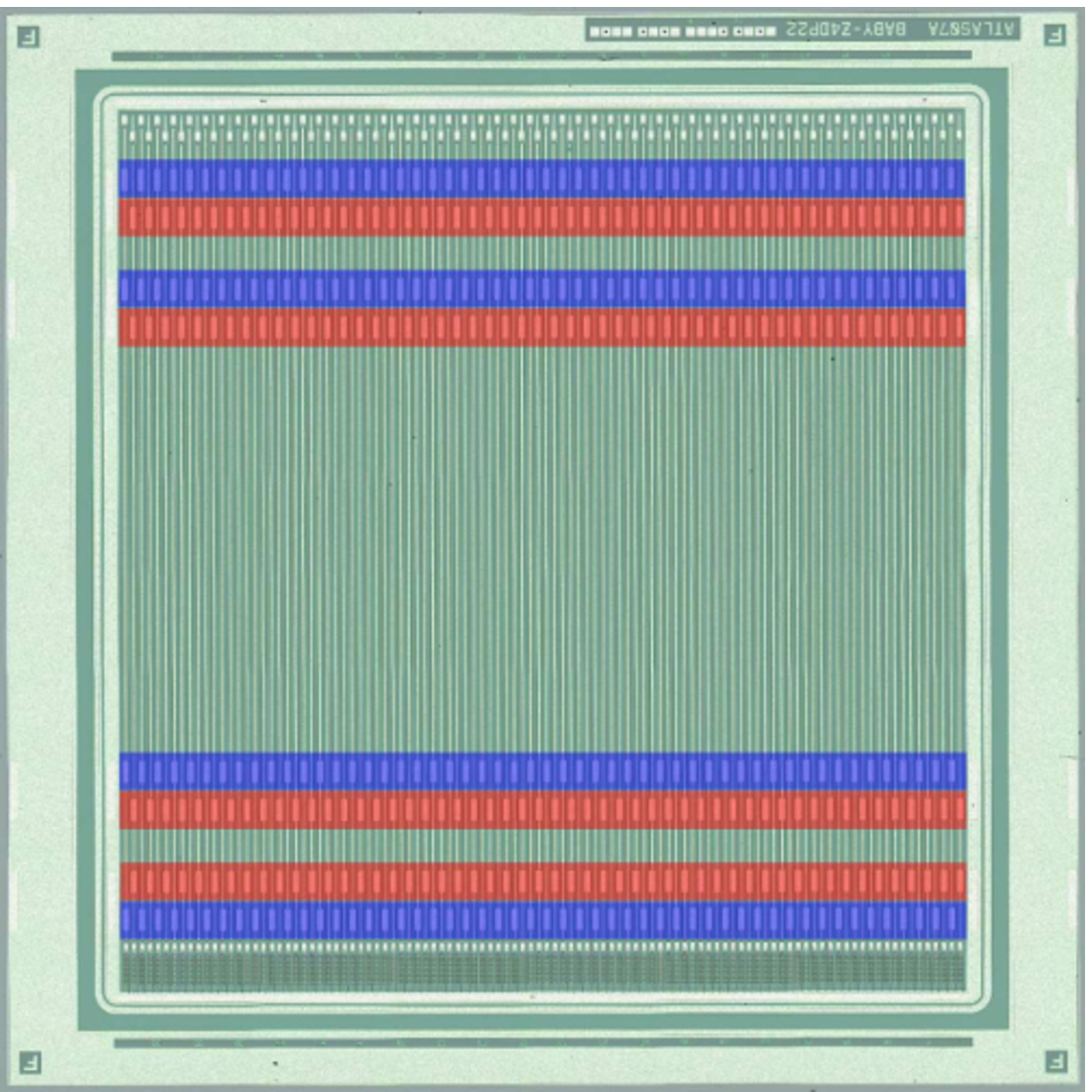}
  \caption{ATLAS07 miniature sensor in the same orientation as used in the test beam. Bond pads of odd and even numbered channels are coloured red and blue, with a total of eight rows of bond pads on the miniature sensor.}
  \label{fig:8_maps_b}
\end{subfigure}
\caption{Sensor hit map in comparison with the sensor layout: the size of the hitmap has been scaled to approximately match the active area of the miniature sensor. The hit map shows that in several areas of the sensor the mainly responding channels are almost exclusively odd or even. The positions of these areas match the bond pad rows indicated on the sensor layout. It can be seen that the pattern of bond pad rows (alternating on even and odd channels in the sensor top half, even-odd-odd-even in the bottom half) is also found in the pattern of mainly responding channels, indicating that the effect found in the hit map is caused by the array of bond pad rows on the sensor.}
\label{fig:8_maps_c}
\end{figure}

The effect was made more visible by dividing the sensor area in a grid with bin sizes of $\unit[14.9\times 149]{\upmu\text{m}^2}$ and finding the sensor channel collecting the most hits for any given position. Figure~\ref{fig:8_maps_a} shows the resulting response map for an ATLAS07 miniature sensor in comparison with its bond pad layout (see figure~\ref{fig:8_maps_b}).
The hit map confirms that the modified sensor architecture in bond pad areas affects the area over which a sensor strip collects charges: a sensor strip responds in a wider area around a sensor bond pad, while neighbour sensor strips respond over a smaller area. 

The effect of sensor strips responding over smaller or wider areas than expected could not be unambiguously attributed to either the presence of bond pads or modified p-stop positions, given the maximum resolution achieved with the telescope in the setup.

\section{Sensor layout considerations}

The ATLAS12~\cite{8_ATLAS12} sensor is divided in four strip segments, where each segment has a length of \unit[$\sim2.5$]{cm} and five rows of bond pads. Each row, consisting of bond pads on odd and even numbered sensor strips, accounts for \unit[700]{$\upmu$m} of modified p-stop positions of which bond pads make up \unit[400]{$\upmu$m}. With overall dimensions of \unit[$9.75\times9.75$]{cm$^2$}, five rows of bond pads on each of the four strip segments lead to a total of \unit[14]{mm} (\unit[14.4]{\%}) of modified p-stops and \unit[8]{mm} (\unit[8.2]{\%}) of bond pads on one sensor strip. 

In these areas, charge collection differs from the expected standard sensor behaviour and thus particle tracking can be affected. Depending on the main contributor to the variations (modified p-stop positions or bond pads), a modification of the sensor layout could be contemplated:
\begin{itemize}
 \item if bond pads were found to affect the responding area of a sensor strip, the number of bond pad rows on the sensor could be reduced
 \item if modified p-stop positions were found to define the area over which a strip responds, the sensor architecture could be modified (using optimised p-stop positions or a sensor architecture with p-spray) 
\end{itemize}
In each case, the implementation of a track reconstruction algorithm including position information associated with the sensitive sensor regions could counteract a negative impact on particle tracking.
In order to identify the mainly defining element of a sensor strip's responding area, a further study with high positioning precision was performed.

\section{Mapping in an X-ray beam}
\label{sec:Dia16}

\begin{figure}
\centering
\includegraphics[width=0.99\textwidth]{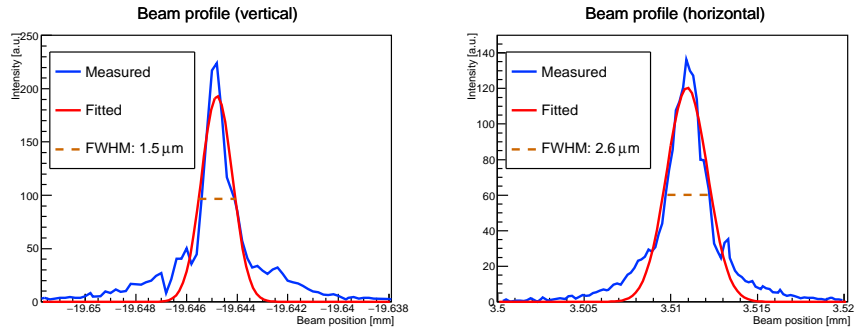}
\caption{Synchrotron X-ray beam profile measured using a gold wire: the horizontal and vertical beam width were measured to be {\unit[1.5]{$\upmu$m}} and {\unit[2.6]{${\upmu}$m}}.}
\label{fig:bp}
\end{figure}

\begin{figure}
\centering
\includegraphics[width=0.6\textwidth]{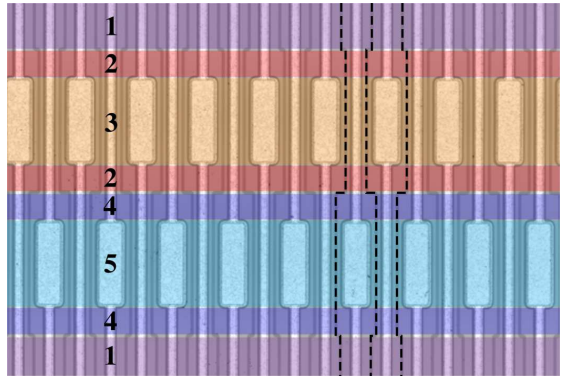}
\caption{Laser microscope image of bond pad rows on sensor showing the different sensor architectures around bond pads. P-stops are visible as dark lines between strip implants with aluminium tops and bond pads. The standard sensor layout with equidistant p-stops is coloured violet (1). Bond pads on odd (light blue, 5) and even (orange, 3) numbered strips show different p-stop positions (dashed black lines), arranged around the bond pads. Between bond pads and standard sensor layout, a transition region (red (2)/dark blue (4)) can be seen, where p-stops are not equidistant, but no bond pads are present.}
\label{fig:8_sensorregions}
\end{figure}

In order to investigate the impact of p-stops and bond pads on the charge collecting area of a sensor strip, a micro-focused \unit[$2\times3$]{$\upmu\text{m}^2$} X-ray beam (see figure~\ref{fig:bp}) was used. 
The sensor was moved in the beam to scan different areas of the sensor (see figure~\ref{fig:8_sensorregions}) in order to compare any potential differences to how sensor strips responded with X-ray focused on various sensor architectures including:
\begin{itemize}
\item equidistant p-stops
\item modified p-stop positions
\item modified p-stop positions around bond pads
\end{itemize}
By using a beam size much smaller than the structures under investigation, differences in the number of collected hits for different sensor areas can be resolved.

\subsection{Setup}
\label{sec51:setup}

Inside the ATLAS ITk, silicon sensors will be operated in a high-radiation environment, thus characteristics of both non-irradiated and irradiated sensors are of interest. Hence, two sensors of the same architecture and nominal active thickness (\unit[300]{$\upmu$m}), one irradiated and one non-irradiated, were tested in comparison:
\begin{itemize}
 \item a non-irradiated ATLAS12~\cite{8_ATLAS12} miniature sensor, attached to an ABC130 readout chip, using the same setup as used in the initial measurements~\cite{8-Diamond1}, where an effect was first observed
 \item one irradiated ATLAS07 miniature sensor, irradiated with reactor neutrons (hardness factor $\kappa = 0.9$~\cite{ref_hardness}), to a fluence of \unit[$2\cdot10^{15}$]{n$_{\text{eq}}$/cm$^2$}~\cite{ref_equivalent}. This sensor was connected to an ALiBaVa readout system, using the same test beam setup as in the DESY test beam (see section~\ref{sec:DESYbeam}).
\end{itemize}
Due to the high irradiation level of the irradiated ATLAS07 sensor, the depletion voltage exceeded the possible bias voltage range of up to \unit[-1000]{V} and could not be determined from measurements. The sensor was thus operated under-depleted at a reverse bias voltage of \unit[-1000]{V} and at a temperature of \unit[-20]{$^{\circ}$C}. The non-irradiated ATLAS12 sensor was operated over-depleted at a bias voltage of \unit[-360]{V} (nominal full depletion voltage: \unit[-300]{V}~\cite{8_ATLAS12}).

The sampling rates of both the ABC130 chips and the ALiBaVa system used for data readout are \unit[25]{ns}. Compared to the distance between two electron bunches in the Diamond Light Source of approximately \unit[2]{ns} (900 bunches distributed over a synchrotron length of \unit[562]{m}), a sampling rate of \unit[25]{ns} contains photons emitted by 12 to 13 bunches. A flux of $\unit[1\cdot10^{8} \pm 20\,\%]{\text{photons}/\text{s}}$ was measured for the applied beam configuration using a calibrated diode, corresponding to an average of $\unit[1.0\pm0.2]{\text{photon}}$ in \unit[10]{ns}.
Using the attenuation coefficient of \unit[15]{keV} photons in silicon of \unit[24.15]{cm$^{-1}$}~\cite{NIST}, the probability of each photon to react within a \unit[300]{$\upmu$m} detector volume is calculated as \unit[51.5]{\%}. The probability for 0, 1, 2 or 3 photons to react with the sensor within a random \unit[25]{ns} readout window can thus be estimated to be \unit[24.2, 36.5, 25 and 10.6]{\%} with an average number of \unit[$1.34\pm0.27$]{photons per event}. 
Due to the long sampling rate compared to the short time between subsequent bunches, data was taken randomly. While multiple photon interactions per sensor resulted in a higher deposited charge, but did not affect the number of collected hits, events with zero interacting photons led to an overall lower number of hits. Hit maps were thus scaled to the highest number of collected hits per bin of the map.

\begin{figure}
\centering
\includegraphics[width=0.7\linewidth]{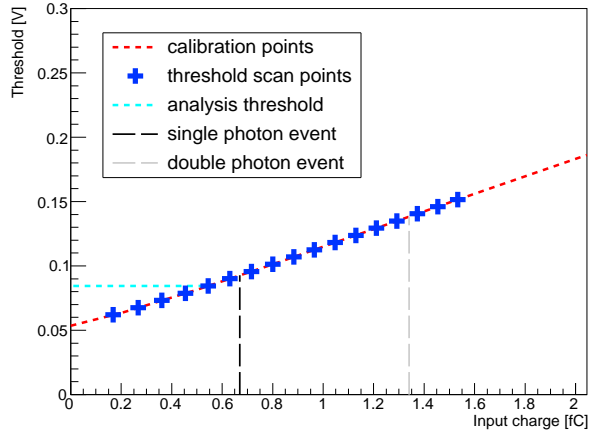}
\caption{Thresholds and input charges for an ABC130 readout chip connected to an ATLAS12 sensor: the thresholds corresponding to any given input charge were calculated using internal calibration circuits of the readout chips. The threshold corresponding to the expected input charge of a {\unit[15]{keV}} electron, {\unit[0.67]{fC}}, was found to correspond to a threshold of {\unit[93]{mV}}. For data taking, a lower readout threshold of {\unit[85]{mV}} was chosen.}
\label{fig:8_ABC130D2017}
\end{figure}

For the irradiated sensor read out by an ALiBaVa system, 100,000 events were collected for each position of the beam on the sensor. Here, clusters in the ALiBaVa system were reconstructed using channels with a signal lying 1 sigma above the determined noise level. Maximal one neighbouring cluster was added if its signal exceeds 1 sigma over the noise level. These thresholds had to be set lower than usual because of the generally lower charge deposited by the X-ray photons and to be sensitive to small signals and small differences between signals. The clustering was restricted to the area hit by the beam (5 strips in total) in order to suppress the creation of random clusters from noise elsewhere in the sensor.
For the non-irradiated sensor attached to an ABC130 readout chip, a threshold scan was performed for thresholds ranging from \unit[62]{mV} to \unit[152]{mV} (see figure~\ref{fig:8_ABC130D2017}). At each beam position, 10,000 triggers for a given threshold were sent to each readout channel. For each trigger, a hit was registered in a channel if its collected charge exceeded the pre-set threshold.

While the different readout systems connected to the two sensors required different modes of data taking concerning the number of triggers and collected hits, the geometrical parameters of scans on both sensors were chosen to be identical (see table~\ref{tab:8_scangeo}).
\begin{table}
\begin{center}
\begin{tabular}{l|c|c}
 & x-direction & y-direction \\
 & (across sensor strips), & (along sensor strips), \\
 & $[\upmu$m$]$ & $[\upmu$m$]$ \\
 \hline
bond pad & 60 & 200 \\
\hline
modified p-stops & & \\
around bond pads & 52 & 350 \\
between bond pads & 97 & 350 \\
\hline
scanning step & 15 & 60 \\
scanning length & 210 & open \\
\end{tabular}
\end{center}
\caption{Parameters of a grid scan over the bond pad area of an ATLAS07 sensor: step sizes were chosen to be smaller than sensor architecture features.}
\label{tab:8_scangeo}
\end{table}
The scan length perpendicular to the strip orientation was chosen to ensure that one strip was entirely covered, including its presumably widest area around its bond pad. Changes of the sensor position with respect to the beam were made using translation stages in x and y-direction, which allowed position changes $\unit[<1]{\upmu\text{m}}$.
It should be noted that while the stages allowed movements with $\unit[<1]{\upmu\text{m}}$ precision, the initial position of the sensor with respect to the beam has to be estimated using a laser alignment system with a positioning precision of about \unit[0.5]{mm}. Hence it was only possible to point the beam next to a bond pad row and move across a region of interest, not to select scan point positions on bond pads or within regions with modified p-stops only (see figure~\ref{fig:8_sensorregions}).
Step sizes were thus chosen to ensure that at least one point of the scanning grid would fall into each of the sensor architectures of interest, in particular the region where p-stops were not equidistant, but no bond pads were present.

\subsection{Results}

\begin{figure}
\centering
\begin{subfigure}{.197\textwidth}
  \centering
  \includegraphics[width=\linewidth]{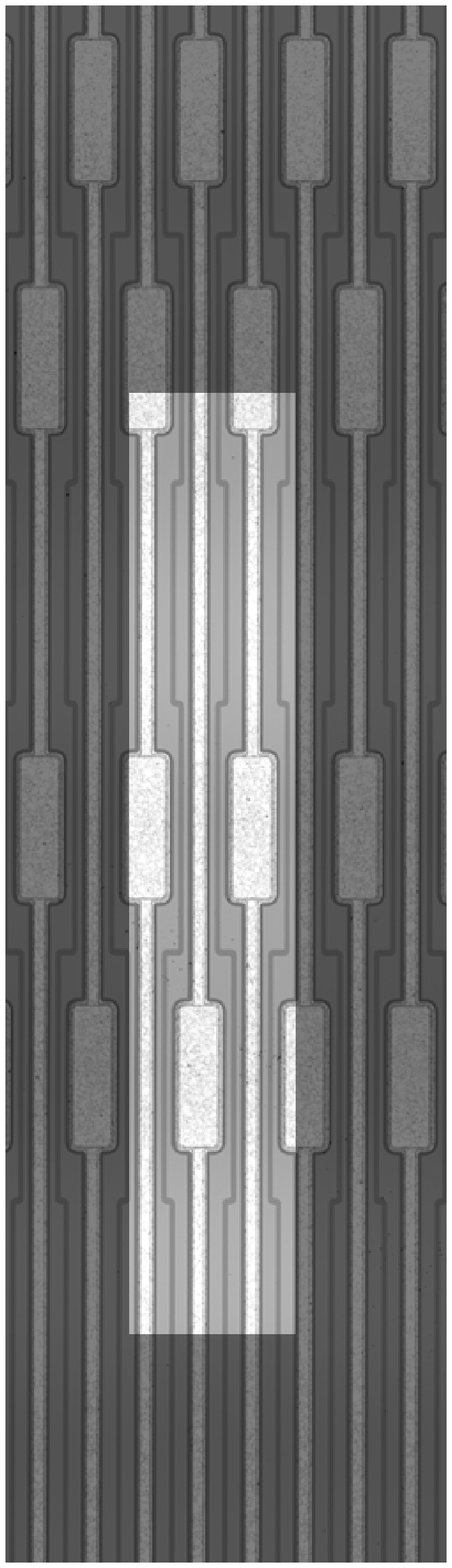}
  \label{fig:8_xmapa}
\end{subfigure}
\begin{subfigure}{.25\textwidth}
  \centering
  \includegraphics[width=\linewidth]{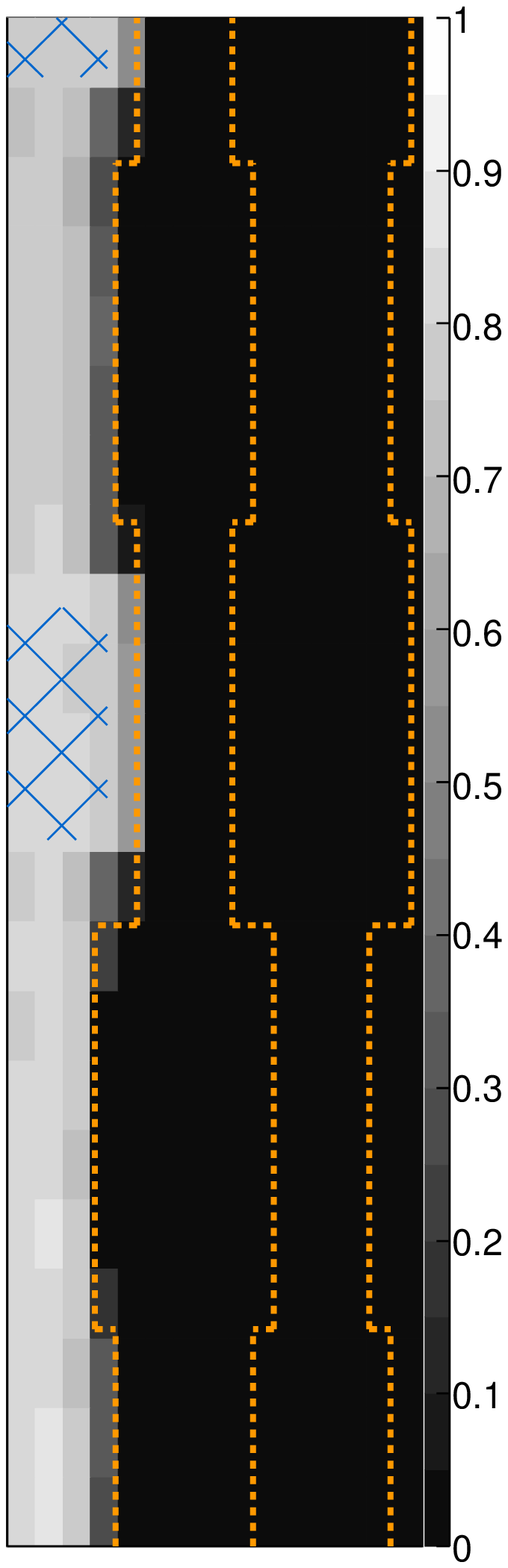}
  \label{fig:8_xmapb}
\end{subfigure}
\begin{subfigure}{.25\textwidth}
  \centering
  \includegraphics[width=\linewidth]{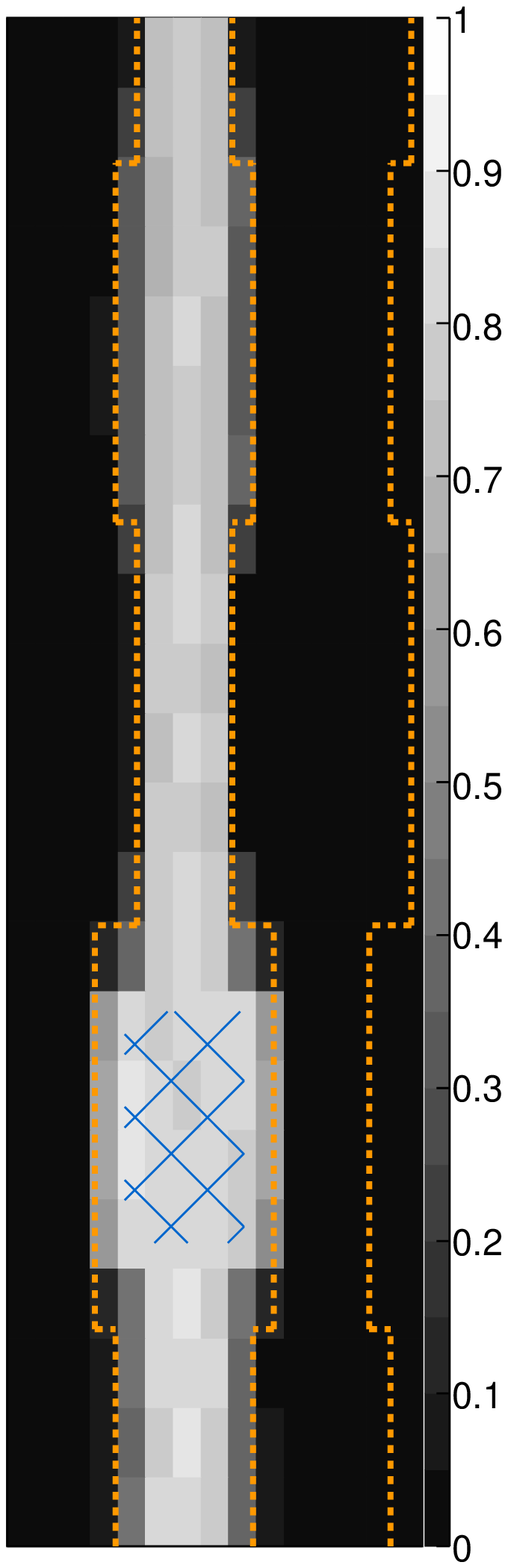}
  \label{fig:8_xmapc}
\end{subfigure}
\begin{subfigure}{.25\textwidth}
  \centering
  \includegraphics[width=\linewidth]{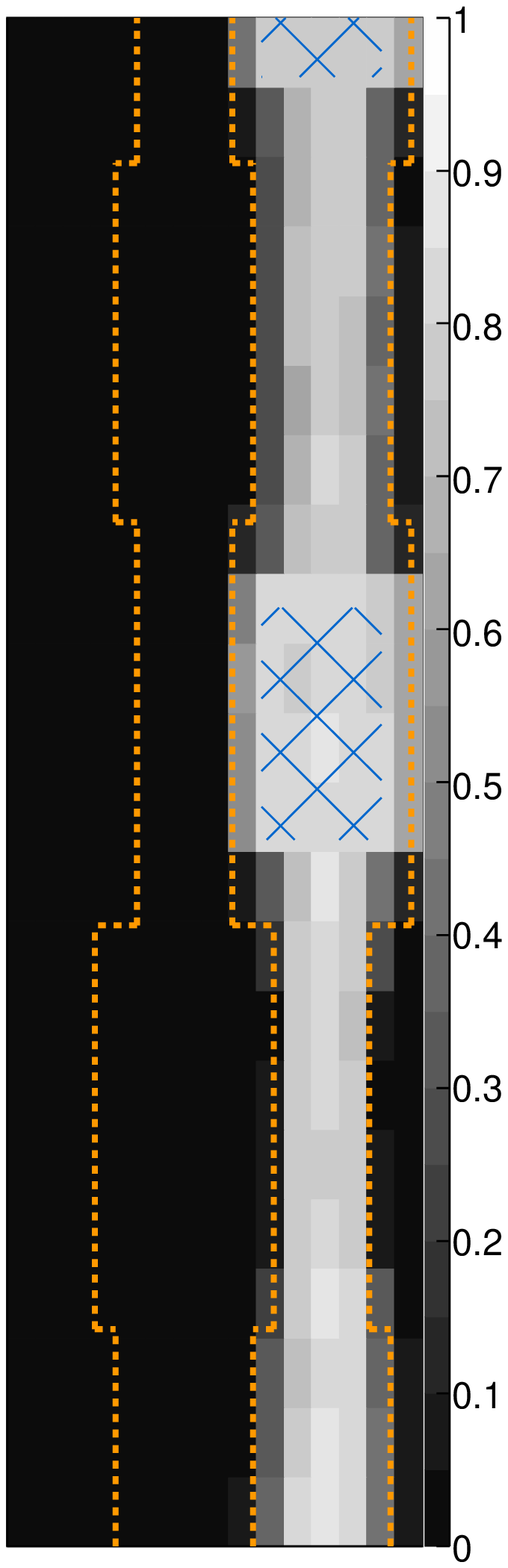}
  \label{fig:8_xmapd}
\end{subfigure}
\caption{Laser microscope image of the approximate sensor area (highlighted) covered in X-ray beam scan and resulting three hit maps. Each plot shows the hits collected by one readout channel for the same scan area of {$\unit[1.26\times0.21]{\text{mm}^2}$}, divided by the maximum number of hits per bin in the map. Positions of p-stops (dashed orange lines) and bond pads (blue shaded areas), determined from fits of collected hit distributions are indicated on the hit maps.}
\label{fig:8_xmaps}
\end{figure}

For each sensor strip covered in the scan, the collected hits for each beam position were plotted to map its responding area. Figure~\ref{fig:8_xmaps} shows the scanned sensor area and the corresponding hit maps.
The results show that the width over which a sensor strip responds does depend on the sensor architecture at that position: the presence of bond pads increases the area over which a sensor responds, with the number of hits collected by neighbour strips decreasing accordingly.

\begin{figure}
\centering
\begin{subfigure}{.3\textwidth}
  \centering
  \includegraphics[width=\linewidth]{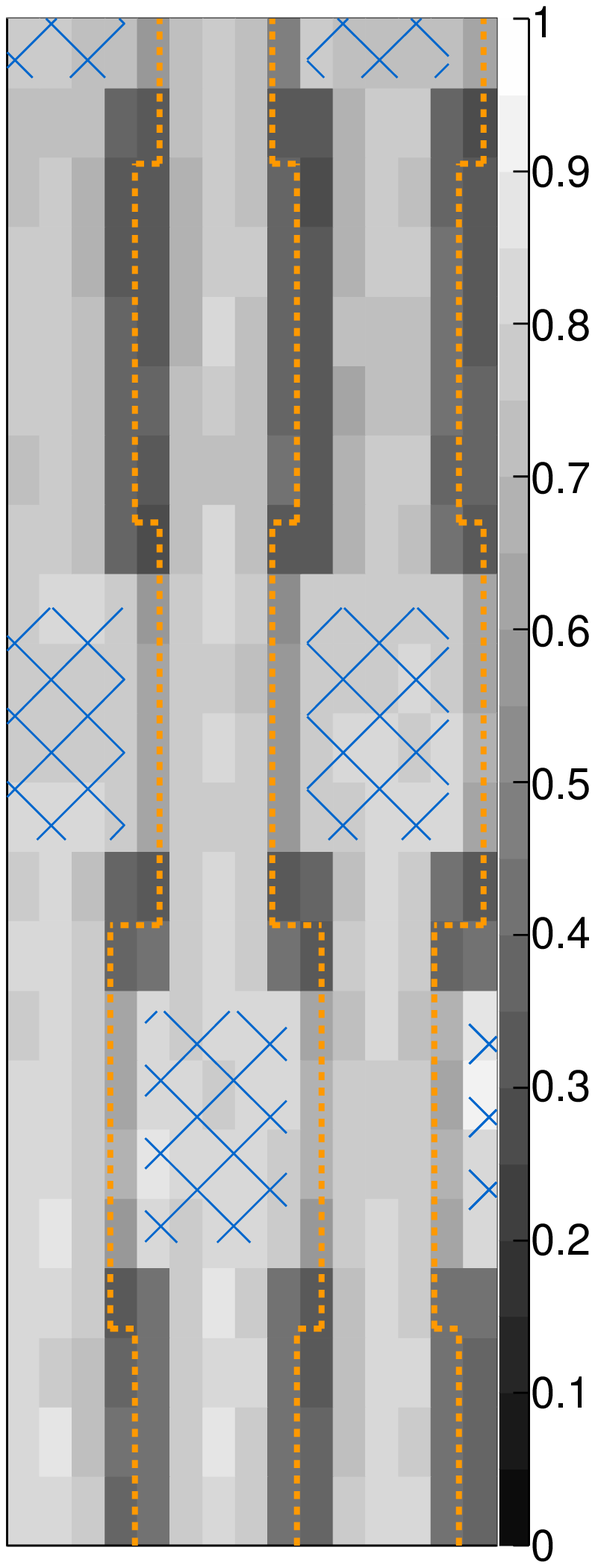}
  \label{fig:8_DAQ4}
  \caption{}
\end{subfigure}
\begin{subfigure}{.3\textwidth}
  \centering
  \includegraphics[width=\linewidth]{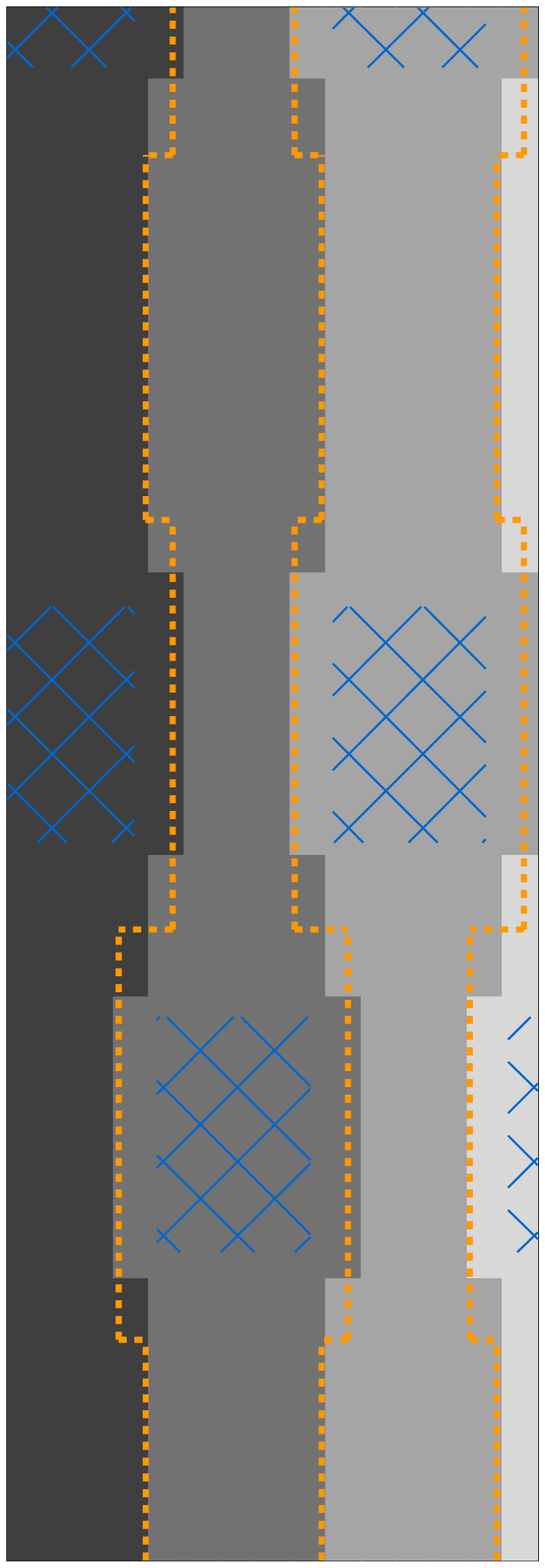}
  \label{fig:8_DAQ5}
  \caption{}
\end{subfigure}
\caption{(a) combined hits relative to the maximum number of hits per bin from four adjacent sensor strips over a {$\unit[1.26\times0.21]{\text{mm}^2}$} area of a sensor bond pad region. Combined hits from neighbour channels show that, around bond pads, the number of collected hits is higher than in the standard sensor area without bond pads. Charge sharing between adjacent strips leads to fewer hits being collected by the binary readout system and a less efficient region between strips. (b) Mapping the channel with the highest number of collected hits for each beam position shows that the area over which a sensor strip responds does not follow the shape of a p-stop: sensor strips show similar responses in areas with equidistant and unevenly spaced p-stops.}
\label{fig:8_mainly}
\end{figure}

In order to investigate the impact of p-stop positions on the responding sensor area, sensor areas without bond pads but with non-uniform p-stop positions were studied. Analogous to the map shown in figure~\ref{fig:8_maps_a}, the number of hits collected in each channel was compared for each bin, showing the highest responding channel for each beam position. Figure~\ref{fig:8_mainly} shows the resulting sensor map.
It was found that strip sensors responding over wider or narrower areas can be attributed mainly to the presence of bond pads, with the p-stop positions having only a minor impact.

\begin{figure}
\centering
\begin{subfigure}{.25\textwidth}
  \centering
  \includegraphics[width=\linewidth]{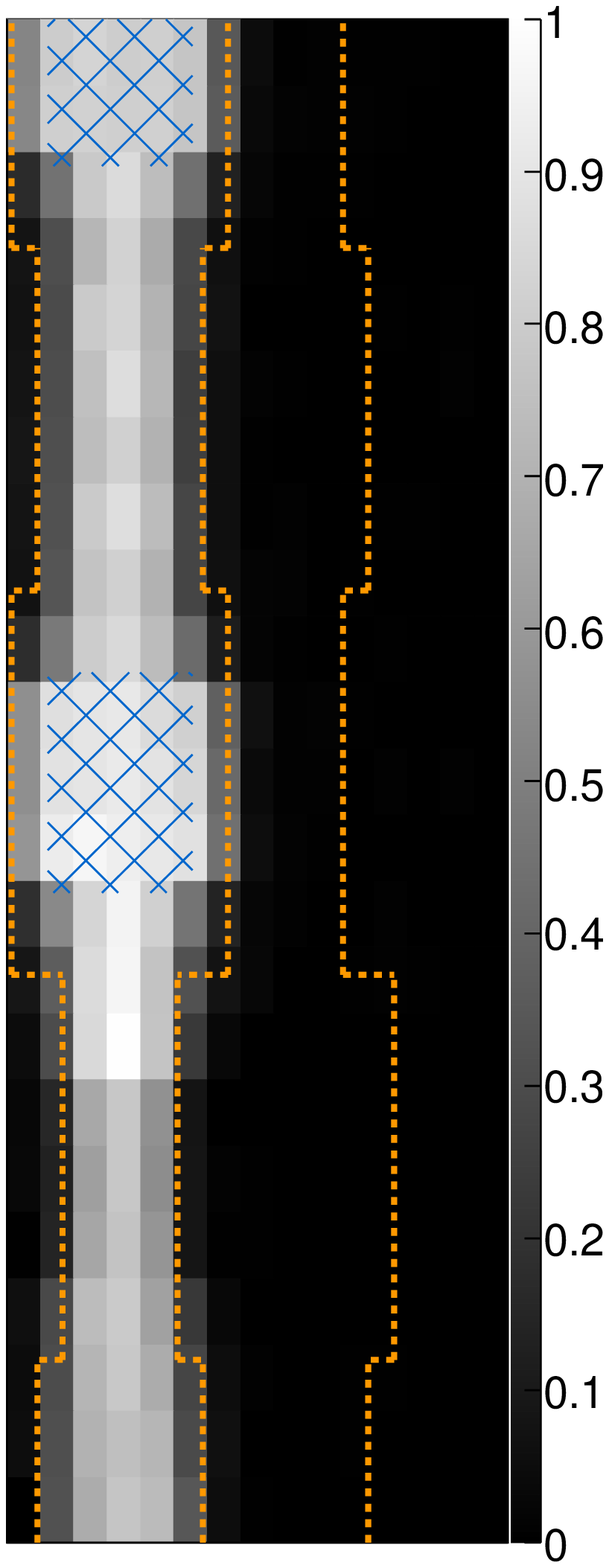}
\end{subfigure}
\begin{subfigure}{.25\textwidth}
  \centering
  \includegraphics[width=\linewidth]{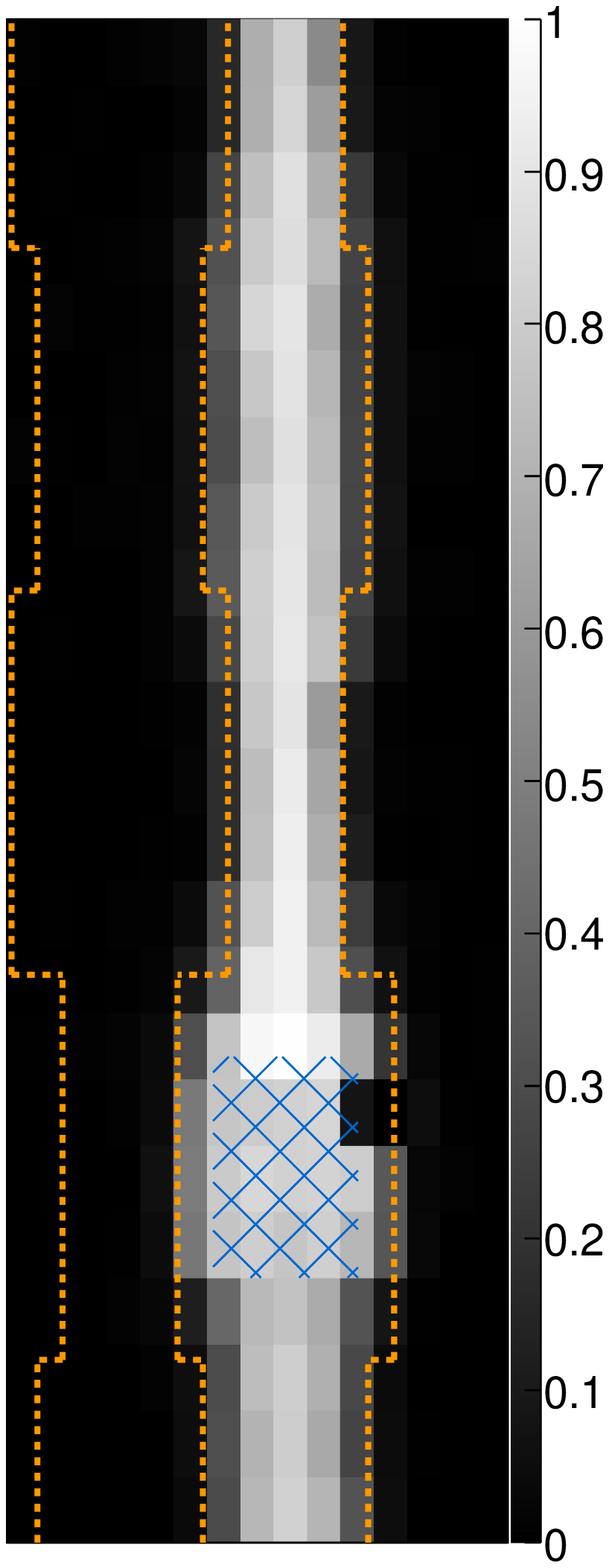}
\end{subfigure}
\begin{subfigure}{.25\textwidth}
  \centering
  \includegraphics[width=\linewidth]{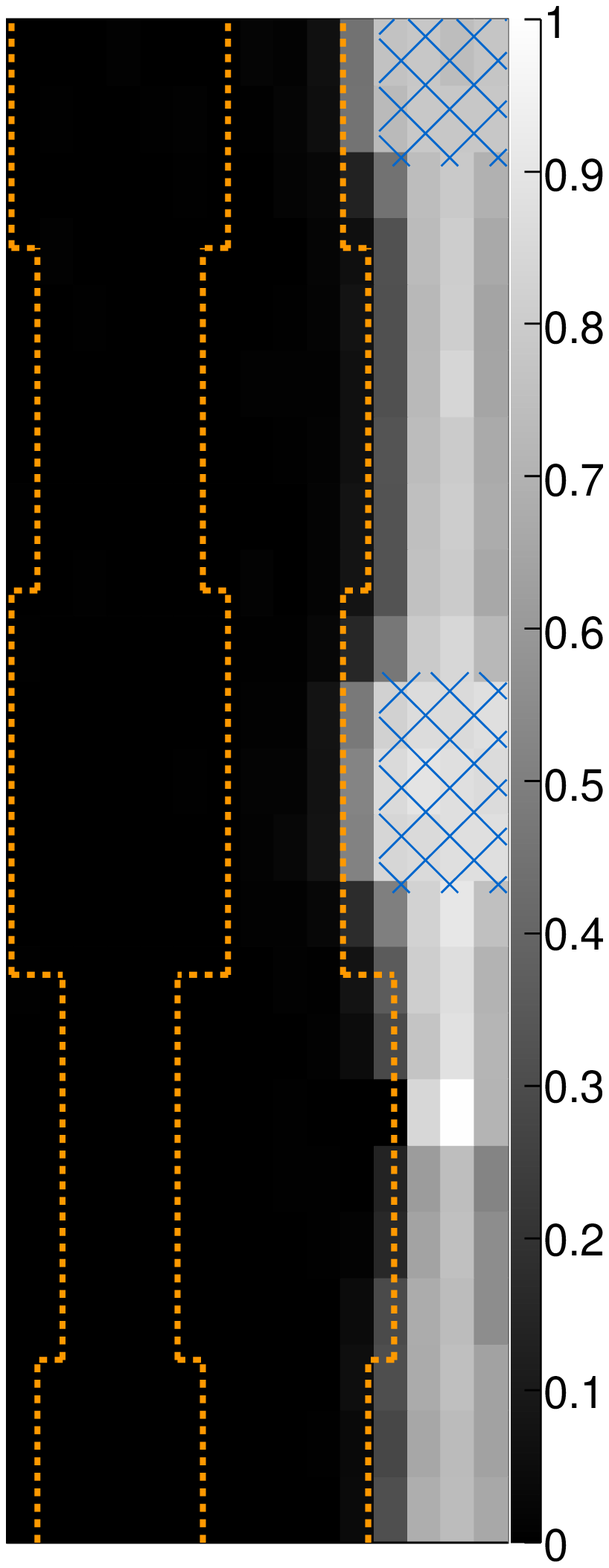}
\end{subfigure}
\caption[Hit maps for three sensor strips over a ${\unit[1.32\times0.21]{\text{mm}^2}}$ sensor grid scan]{Hit maps for three sensor strips over an area of ${\unit[1.32\times0.21]{\text{mm}^2}}$ in steps of ${\unit[60\times15]{\upmu\text{m}^2}}$. Hit numbers are shown relative to the maximum number of hits collected for one sensor strip. Positions of sensor features (p-stops (dashed orange lines) and bond pads (shaded blue areas) were determined from hit distributions and are shown on the maps. While the responding areas of individual strips are larger around bond pads, no difference could be observed between areas with equidistant p-stops and unevenly spaced p-stops.}
\label{fig:8_map_irrad}
\end{figure}

Irradiation influences the electric field of the sensor and thus the responding area of each sensor strip~\cite{ref_radiation}. The results obtained for a non-irradiated sensor were compared to a similar scan performed to an ATLAS07 sensor irradiated to a fluence of $\unit[2\cdot10^{15}]{\text{\unit[1]{MeV neutrons}}/\text{cm}^2}$ using reactor neutrons. This corresponds to the full High Luminosity LHC dose expected in the ATLAS ITk strip detector, including a safety factor of 2. Figure~\ref{fig:8_map_irrad} shows the hit maps obtained from individual sensor strips of an irradiated sensor. Due to limited available beam time, the scanned area on the irradiated sensor was smaller than the area scanned on the non-irradiated sensor.
Similar to the non-irradiated sensor, hit maps for sensor strips on an irradiated sensor show increased numbers of hits around bond pads. 
While scan steps had been chosen to contain at least one row of scanned points within the area of modified p-stops without bond pads, the initial beam position (see section~\ref{sec51:setup}) was found to lead to scan points being located mostly in areas at least partially covered by bond pads or mixed with standard p-stop positions. Only one row of scan points (located below the first bond pad, see figure~\ref{fig:8_map_irrad}) was contained entirely within a region of modified p-stops only where the collected hits can be seen to show only minor changes for standard and altered p-stop positions. Analogous to the findings for a non-irradiated sensor, hitmaps for an irradiated sensor hence indicated that changes in the number of collected hits are mostly caused by the presence of bond pads, with altered p-stop positions having only a minor impact.

\begin{figure}
\centering
\includegraphics[width=0.7\textwidth]{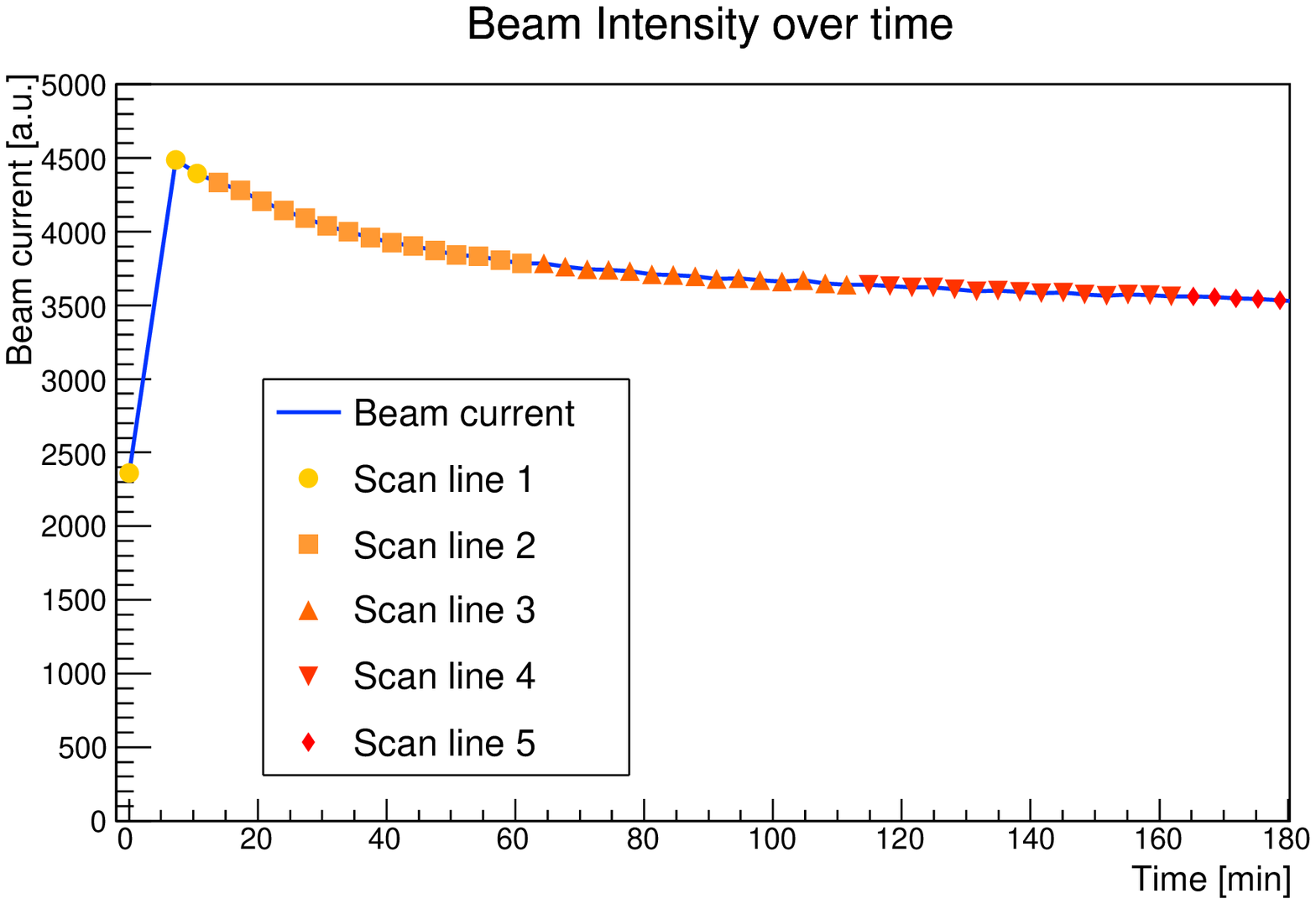}
\caption{Synchroton beam line current measured after a restart as a function of time. Markers represent one beam position of the scan grid, with one scan line consisting of 15 scan points across the sensor strips.}
\label{fig:8_beamintensity}
\end{figure}

It should be noted that the number of photons passing through the sensor for each beam position was found to vary over time, translating into different numbers of hits being collected. Figure~\ref{fig:8_map_irrad} shows a visible discrepancy in collected hits between two areas on the hit map, with the transition being marked by one bin showing significantly fewer entries than the surrounding positions.
Comparing the timestamps of each beam position with the beam current, changes over time were found to match the variations observed in the numbers of collected hits. Figure~\ref{fig:8_beamintensity} shows the measured beam current over time.
The hit map entry with the low number of entries was found to correspond to a beam loss, leading to a low number of photons and registered hits. After restarting, the beam current was higher than before and slowly decreasing, translating into fewer hits collected by the sensor. The higher beam intensity after restart led to the number of corresponding hits increasing by \unit[25]{\%}, with the subsequent \unit[22]{\%}-beam intensity decrease (see figure~\ref{fig:8_beamintensity}) translating into hit numbers decreasing by \unit[11]{\%}.
While the changes of the beam current complicate absolute statements about collected hits and efficiency, the variations were small enough to allow for the comparison of responding sensor areas.

\section{Comparison with sensor simulations}

\begin{figure}
\centering
\begin{subfigure}{.49\textwidth}
  \centering
  \includegraphics[width=\linewidth]{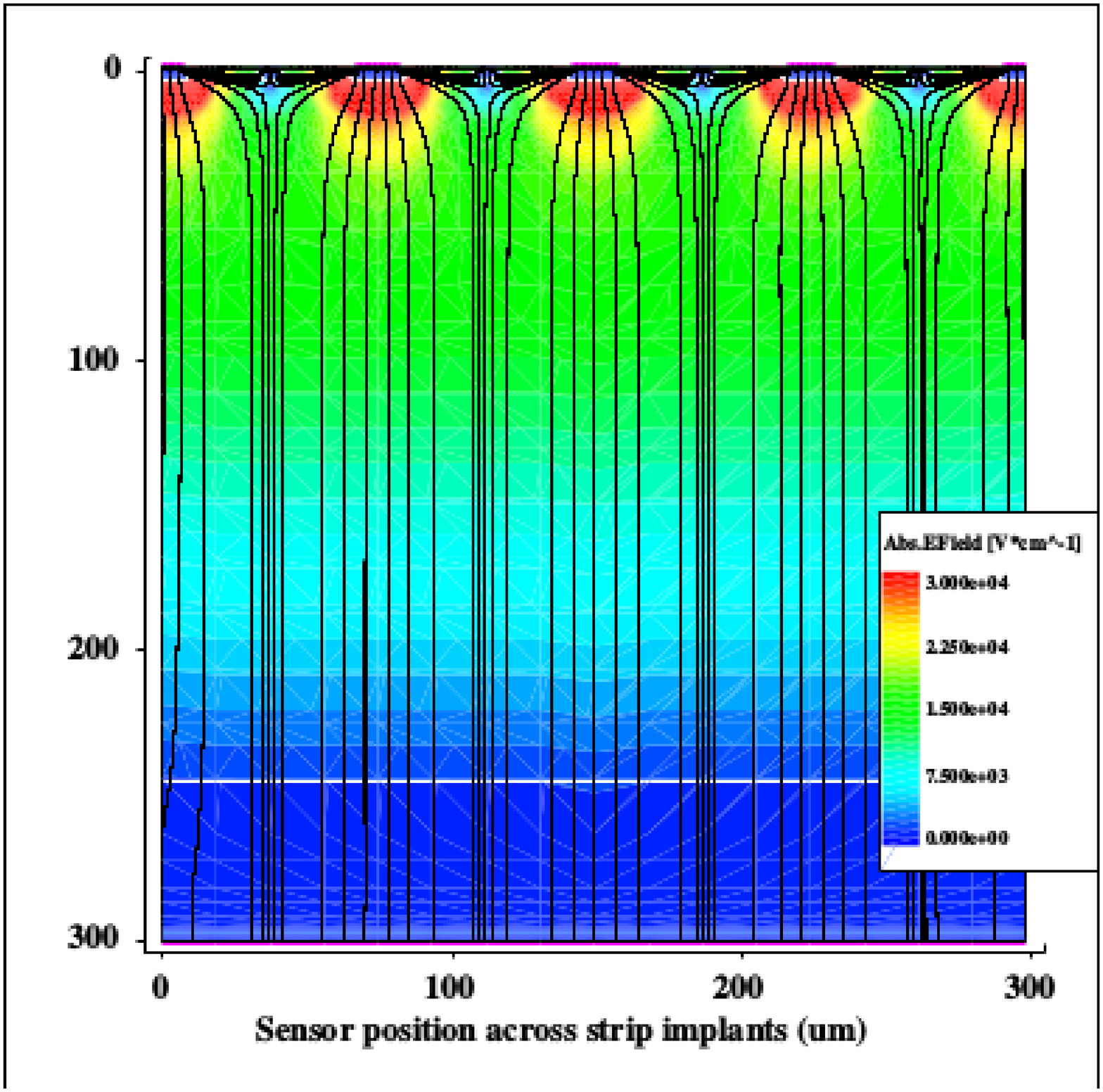}
  \caption{Simulation of the standard sensor layout with equidistant p-stop positions}
  \label{fig:sims_1}
\end{subfigure}
\begin{subfigure}{.49\textwidth}
  \centering
  \includegraphics[width=\linewidth]{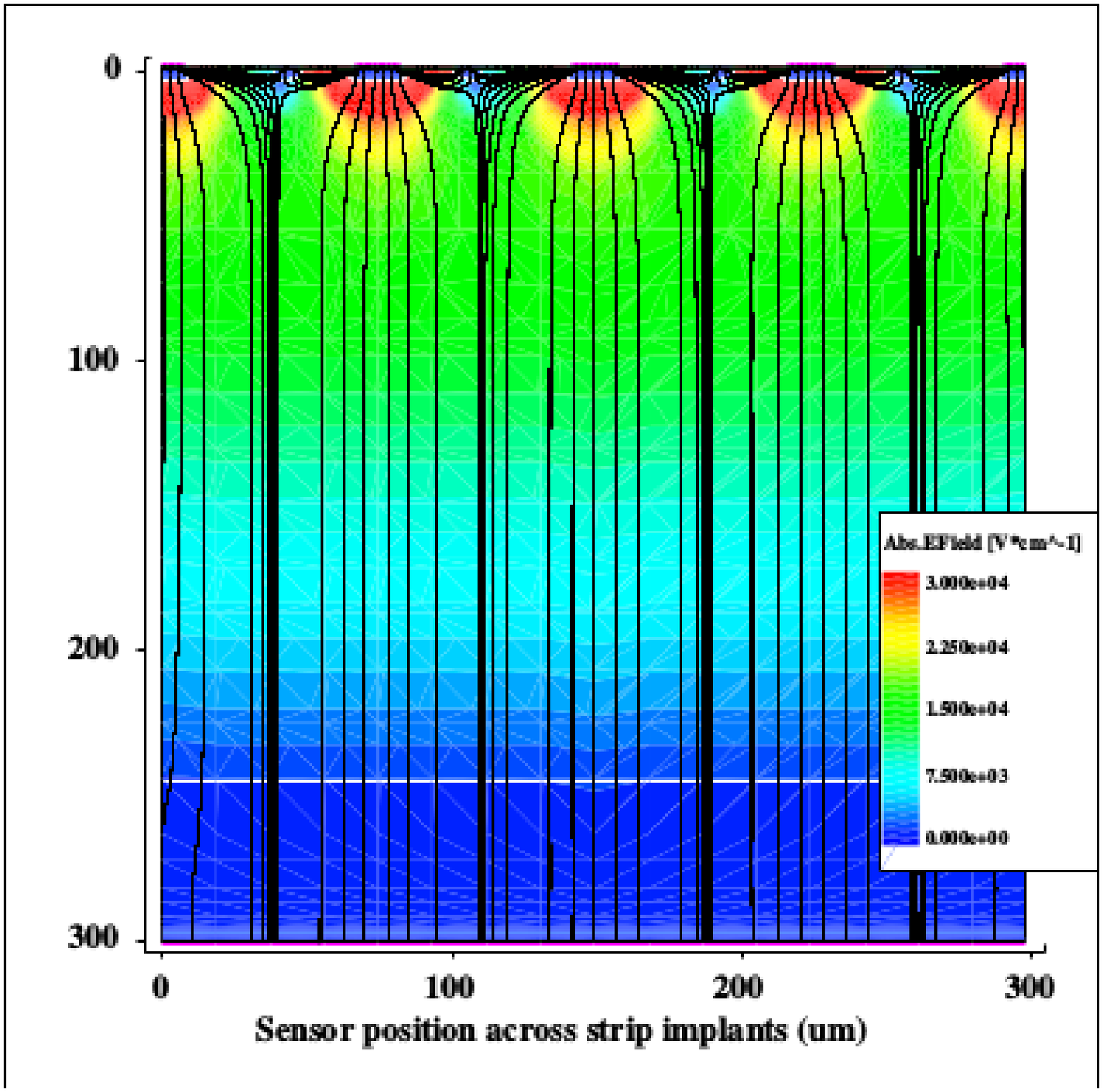}
  \caption{Simulation of the sensor layout with altered p-stop positions}
  \label{fig:sims_2}
\end{subfigure}
\begin{subfigure}{.49\textwidth}
  \centering
  \includegraphics[width=\linewidth]{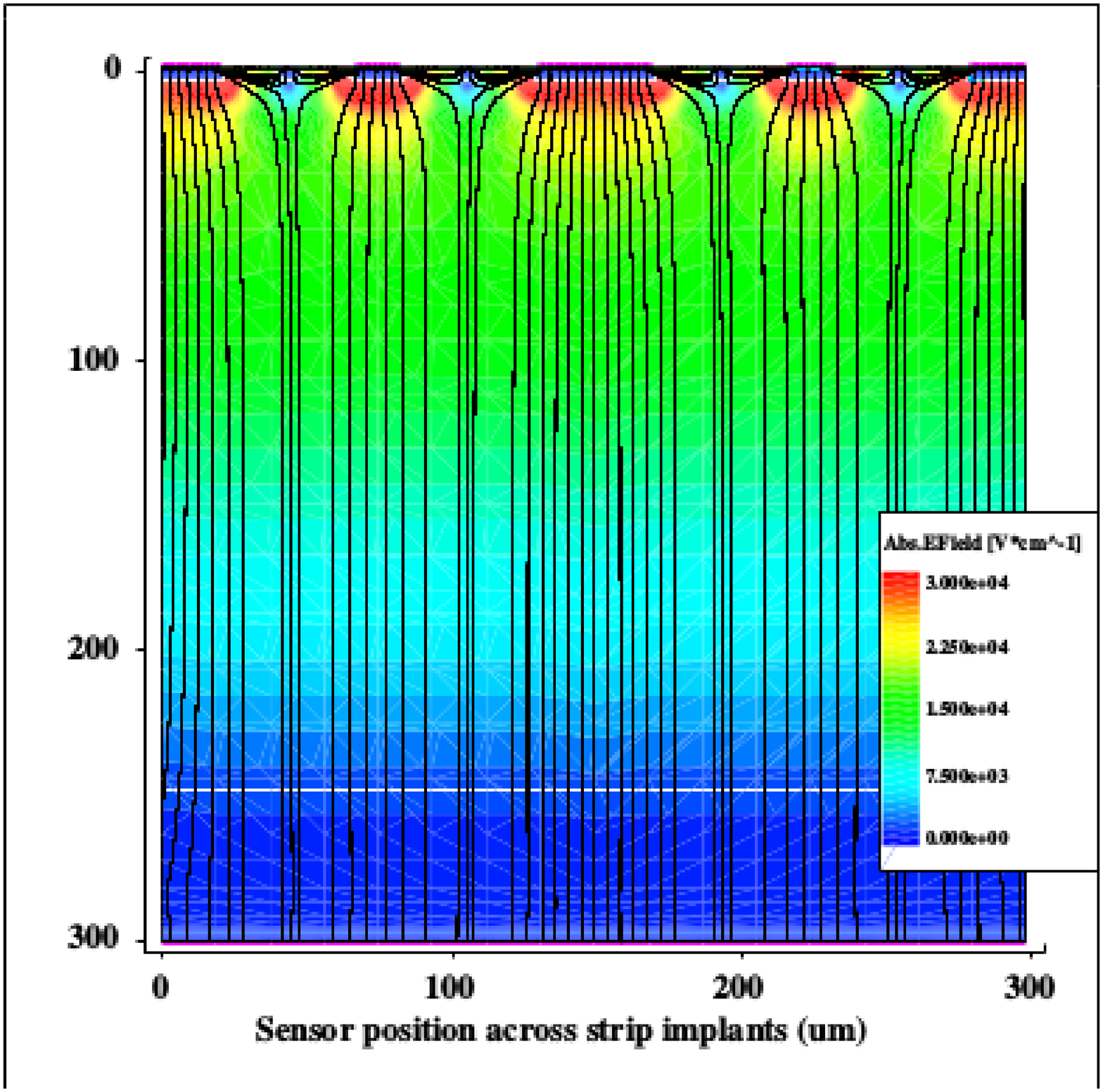}
  \caption{Simulation of the sensor layout with bond pads and altered p-stop positions}
  \label{fig:sims_3}
\end{subfigure}
\caption{2D TCAD simulation of the electric field and its streamlines of a silicon strip sensor. The simulated cross-section of the sensor shows a width and thickness of {\unit[300]{$\upmu$m}}. A reverse bias of {\unit[-300]{V}} is applied at the back side of the sensor at {\unit[300]{$\upmu$m}}. The simulations are done for a resistivity of {\unit[2.5]{k$\Omega\cdot$cm}}, to study the effect of different widths of the metal contact and strip implant, and the effect of different p-stop positions.}
\label{fig:sims}
\end{figure}

The observed charge collection behaviour of the sensors was compared to 2D TCAD simulations of ATLAS07/ATLAS12 sensor architectures~\ref{fig:sims}. Layout modifications matching the different sensor regions showed that the position of p-stops around strip implants has only a minor impact on the electric field inside a sensor (see figures~\ref{fig:sims_1} and~\ref{fig:sims_2}), matching the observations found in test beam measurements. 

Areas between strip implants show a lower electric field strength than areas below strip implants, which agrees with the hit maps obtained in test beams showing fewer hits being collected at the edges of a strip than in its centre. The presence of a wider implant and bond pad (see figure~\ref{fig:sims_3}) was found to extend the region of higher electric field strength further towards the edges of a strip, leading to a more homogeneous field around the implant. A larger area of high field strength can be assumed to lead to better charge collection in the corresponding sensor region.
The simulations agreed well with the wider areas of collected charge below bond pads and minor impact of p-stop positions observed in test beams.
After irradiation, local defects disturb the electric field of a sensor~\cite{ref_trapping}, reducing the overall charge collecting behaviour, particularly at the less efficient edges of sensor strips. Similar to the behaviour observed for a non-irradiated sensor, the wider implant and added bond pads in bond regions were found to increase the increase the charge collecting area of a sensor strip towards its edges. A similar effect has been observed in TCT measurements of irradiated sensors~\cite{ref_TCT}.
The altered sensor architecture in bond pad regions can thus be assumed to have a similar beneficial effect on the electric field of an irradiated sensor as simulated for a non-irradiated sensor.

\section{Conclusion and Outlook}

Studies of the bond pad regions of silicon strip sensors with high spatial resolution have confirmed that on ATLAS07 and ATLAS12 sensors, strips respond over a larger width when bond pads are present. It was found that the strip response can be attributed mainly to the geometry of bond pads, with the impact of p-stop positions being much smaller. Similar effects were observed in 2D TCAD simulations of comparable layout alterations.

Bond pads were found to increase the local width over which a sensor strip collects hits from \unit[74.5]{$\upmu$m} to $\unit[\sim95]{\upmu\text{m}}$ and reduce the responding width of adjacent strips to $\unit[\sim54]{\upmu\text{m}}$, leading to corresponding variations in the numbers of collected clusters. Comparable effects can be assumed to occur in sensors with similar architecture features.

Detector simulation studies will be conducted in order to investigate the potential impact of effective strip widths varying along the strip length on the tracking performance. Significant negative impacts on the tracking performance could be the base for modifications of the final sensor layout.

\section*{Acknowledgements}

The measurements leading to these results have been performed at the Test Beam Facility at DESY Hamburg (Germany), a member of the Helmholtz Association (HGF). 
We thank the test beam coordinators and telescope support, Dr. Marcel Stanitzki and Dr. Jan Dreyling-Eschweiler in particular, for their help and support during the test beam.

We thank Diamond Light Source for access to beam line B16 (proposal number MT13500) that contributed to the results presented here. The authors would like to thank personnel of the B16 beam, especially Andy Malandain, for providing advice, support and maintenance during the experiment.

This project has received funding from the European Union’s Horizon 2020 Research and Innovation programme under Grant Agreement no. 654168. The authors would like to thank Dr. Vladimir Cindro from JSI for his help with the sensor irradiations.

The work at SCIPP was supported by the Department of Energy, grant DEFG02-13ER41983.

\bibliographystyle{unsrt}
\bibliography{bibliography.bib}

\end{document}